\documentclass[useAMS,usenatbib]{mn2e}
\usepackage{natbib}
\usepackage{graphicx}
\usepackage{subfigure}
\usepackage{multirow}
\usepackage{amsmath}
\usepackage{amssymb}
\usepackage{epstopdf}
\usepackage{rotating}
\usepackage{soul}
\usepackage{placeins}

\usepackage{color}
\usepackage{hyperref}

\title[High-z SMBH mergers in simulations with dynamical friction]{High-redshift supermassive black hole mergers in simulations with dynamical friction modelling}

\author[DeGraf et al.]{Colin DeGraf$^{1}$, Nianyi Chen$^{2}$, Yueying Ni$^{3}$, Tiziana Di Matteo$^{2,4,5}$,  \newauthor Simeon Bird$^6$, Michael Tremmel$^7$, Rupert Croft$^{2,4}$ \\
$^{1}$ Department of Physics, Truman State University, Kirksville, MO 63501, USA \\
$^{2}$ McWilliams Center for Cosmology, Department of Physics, Carnegie Mellon University, Pittsburgh, PA 15213 \\
$^{3}$ Center for Astrophysics, Harvard \& Smithsonian, Cambridge, MA 02138, US \\
$^{4}$ NSF AI Planning Institute for Physics of the Future, 
Carnegie Mellon  University, Pittsburgh, PA 15213, USA \\
$^{5}$ OzGrav-Melbourne, Australian Research Council Centre of Excellence for Gravitational Wave Discovery\\
$^{6}$ University of California, Riverside,  Riverside CA 92521, USA\\
$^{7}$ Physics Department, University College Cork, Cork, Ireland\\
}

\begin{document}

\maketitle
\begin{abstract}
    In the near future, projects like LISA and Pulsar Timing Arrays are expected to detect gravitational waves from mergers between supermassive black holes, and it is crucial to precisely model the underlying merger populations now to maximize what we can learn from this new data.  Here we characterize expected high-redshift ($z > 2$) black hole mergers using the very large volume {\small{Astrid}} cosmological simulation, which uses a range of seed masses to probe down to low-mass BHs, and directly incorporates dynamical friction so as to accurately model the dynamical processes which bring black holes to the galaxy center where binary formation and coalescence will occur. The black hole populations in {\small{Astrid}} include black holes down to $\sim 10^{4.5} M_\odot$, and remain broadly consistent with the TNG simulations at scales $> 10^6 M_\odot$ (the seed mass used in TNG).  By resolving lower-mass black holes, the overall merger rate is $\sim 5\times$ higher than in TNG.  However, incorporating dynamical friction delays mergers compared to a recentering scheme, reducing the high-z merger rate mass-matched mergers by a factor of $\sim 2 \times$. We also calculate the expected LISA Signal-to-Noise values, and show that the distribution peaks at high SNR ($>$100), emphasizing the importance of implementing a seed mass well below LISA's peak sensitivity ($\sim 10^6 M_\odot$) to resolve the majority of LISA's GW detections.   
\end{abstract}

\section{Introduction}

Supermassive black holes (SMBHs) have been found to exist at the center of galaxies \citep{KormendyRichstone1995}, with a strong correlation to host galaxy properties \citep{Magorrian1998, Gebhardt2000, Graham2001, Ferrarese2002, Tremaine2002, HaringRix2004, Gultekin2009, McConnellMa2013, KormendyHo2013, Reines2015, Greene2016, Schutte2019}.  These correlations hold true across cosmic time, suggesting a coevolutionary growth between black holes and the galaxies which host them.  These galaxies (and the dark matter halos in which they are located) are expected to merge \citep[e.g.][]{Fakhouri2010, Rodriguez-Gomez2015}.  Following a galaxy merger, the central black hole from each progenitor galaxy can migrate toward the galactic center of the newly merged galaxy, where they can form a binary and eventually merge together themselves \citep[e.g.][]{Mayer2007}. 
Black hole mergers produce strong gravitational wave (GW) signals, and the coalescence of a pair of supermassive black holes found at the center of galaxies will produce the strongest GW signals in the Universe.

In the past several years, gravitational waves produced by black hole mergers have been detected using interferometers \citep[e.g.][]{Abbott2016}, but size limitations due to the ground-based nature of these instruments mean that detections to this point have been limited to mergers between stellar mass black holes.  Higher-mass mergers (i.e. between SMBHs) produce GWs with much longer wavelengths, beyond the sensitivity of ground-based interferometers.  However, the upcoming Laser Interferometer Space Antenna (LISA) space mission will focus on lower-frequency GWs corresponding to higher-mass mergers, with sensitivities peaking at $\sim 10^4-10^7 M_\odot$ \citep{AmaroSeoane2017}.  Furthermore, Pulsar Timing Arrays should be capable of detecting even higher mass mergers, reaching black holes above $10^8 M_\odot$ \citep[e.g.][]{Verbiest2016, Desvignes2016, Reardon2016, Arzoumanian2018}.  The GWs detected by these observations should provide a new and powerful mechanism to study SMBHs and their connection with their host galaxies. 

SMBH mergers detected through GWs can provide a wide range of constraints on our understanding of black hole - galaxy coevolution, including estimating the rate at which SMBHs merge \citep[e.g.][]{Klein2016, Salcido2016, Kelley2017, Ricarte2018, Katz2020, Volonteri2020}, the expected merger/coalescence timescale \citep[e.g.][]{Volonteri2020, Banks2022}, how mergers influence the scaling relation between black holes and their host galaxies \citep[e.g.][]{VolonteriNatarajan2009, Simon2016, Shankar2016}, gas environment and accretion efficiencies \citep[e.g.][]{Kocsis2011, Barausse2014, Derdzinski2019}, how SMBH seeds initially form \citep[e.g.][]{Sesana2007, Ricarte2018, DeGraf2021Morphology}, and the potential connection with host galaxy morphologies \citep[with multimessenger studies that combine GW and electromagnetic information, e.g.][]{Volonteri2020, DeGraf2021Morphology}.  To maximize what we can learn from the initial SMBH merger detections, it is important to understand the underlying merging populations.

Cosmological simulations provide an ideal mechanism to characterise merging populations, as they self-consistently model both black holes and galaxy formation, encompass large volumes which provide robust statistical samples, and span a wide redshift range to investigate evolution over cosmic time.  Current simulations \citep[e.g.][]{Vogelsberger2014a, Dubois2014, Schaye2015, Feng2016, Pillepich2018b, Henden2018, Dave2019, Chen2022} resolve a wide range of scales, with BHs ranging from $\sim 10^4-10^{10} M_\odot$, resolving black hole growth, mergers, and their host galaxy properties.  The majority of cosmological simulations, however, tend to only resolve higher-mass black holes (on the order of $10^6 M_\odot$), which misses the majority of LISA-detectable mergers.  Additionally, simulations frequently do not physically model the infall/inspiral of the black holes leading to coalescence, which has the potential to significantly impact merging black hole populations \citep[see, e.g.][]{Volonteri2020, Banks2022}.  Here we study the {\small{Astrid}} simulation \citep{Ni2022}, which includes both low-mass seeds (down to $10^{4.5} M_\odot$), thereby resolving mergers at the peak of LISA's sensitivity; and it directly incorporates dynamical friction to model the orbital dynamics to small scales, providing a more accurate probe of inspiralling black holes, and prevents mergers from occurring during fly-by encounters \citep{Chen2022Astrid}.

This paper is organized as follows: in Section \ref{sec:method} we provide an overview of the {\small{Astrid}} simulation, including black hole and dynamical friction models.  In Section \ref{sec:merger_populations} we discuss the overall black hole population in {\small{Astrid}}, as well as the expected merging populations.  In Section \ref{sec:GWs} we investigate the gravitational waves emitted by these mergers, and combine with LISA sensitivity to calculate the expected Signal-to-Noise (SNR) ratios for the full merger samples (Section \ref{sec:SNR}).  Finally, we summarize our conclusions in Section \ref{sec:conclusions}.  

\section{Method}
\label{sec:method}

In this work, we use {\small{Astrid}} \citep{Bird2022}, a cosmological simulation run using a version of the \small{MP-Gadget} smoothed-particle hydrodynamics (SPH) simulation code, a highly scalable version of the \small{Gadget-3} code \citep{Springel2005}.  The simulation consists of a (250 $h^{-1}$ Mpc )$^3$ volume containing $2\times 5500^3$ particles, resolving galactic halos down to $10^9 M_\odot$ from $z=99$ to $z=2$.  The cosmological parameters for the simulation are based on measurements from \citet{PlanckCosmological2020} ($\Omega_0 = 0.3089, \Omega_\Lambda = 0.6911, \Omega_b = 0.0486, \sigma_8 = 0.82$, and $h=0.6774$), and has an initial mass resolution of $M_{DM}=6.74 \times 10^6 \ h^{-1} \ M_\odot , M_{gas}=1.27 \times 10^6 \ h^{-1} \ M_\odot$ and gravitational softening length of $\epsilon = 1.5 \ h^{-1} \ kpc$.  

The {\small{Astrid}} simulation includes detailed models for galaxy formation and evolution, including reionization \citep{Battaglia2013, FaucherGiguere2020} with self-shielding \citep{Rahmati2013}, star formation \citep{SpringelHernquist2003} with associated feedback \citep{Okamoto2010} and metal return \citep{Vogelsberger2013, Pillepich2018}.  For a more detailed description of the physics modelled in this simulation, see \citet{Ni2022, Bird2022}.

Of particular import for this analysis is the implementation of black holes in {\small{Astrid}} \citep{Ni2022, Chen2022Astrid}.  Black holes in the {\small{Astrid}} simulation are treated as collisionless sink particles, inserted into halos above a mass threshold of $M_{\rm{halo}}=5 \times 10^9 \ h^{-1} \ M_\odot$ and $M_* = 2 \times 10^6 \ h^{-1} \ M_\odot$ which do not already contain a black hole particle.  Rather than using a fixed seed mass for black holes, {\small{Astrid}} selects a mass for each newly-seeded black hole from a power law distribution (power law index $= -1$) from $3 \times 10^4 - 3 \times 10^5 \ h^{-1} \ M_\odot$, intended to remain broadly consistent with a variety of SMBH formation pathways and their subsequent growth \citep[e.g.][]{BegelmanRees1978, MadauRees2001, Volonteri2003, BrommLoeb2003, ReganHaehnelt2009, Katz2015, DeGrafSijacki2020}.  Once seeded, the black holes grow by merging with other black holes, and via mass accretion following a model based on a \citet{BondiHoyle1944}-like formalism applied to the SPH kernel of the black hole (periods of super-Eddington accretion are permitted, but capped at $2 \times$ the Eddington rate). Accreting black holes are assumed to produce a bolometric luminosity proportional to the accretion rate (at 10\% efficiency), and 5\% of the radiated energy is assumed to couple thermally to the surrounding gas (this feedback energy is deposited isotropically among gas particles within the SPH kernel).  Please see \citet{Ni2022} for additional details.

Rather than using a respositioning scheme to simply move all black holes toward nearby potential minima, {\small{Astrid}} implements a dynamical friction model for black holes \citep{Tremmel2015, Chen2022}.  We assume a Maxwellian distribution for the velocity distribution of the surrounding particles (both stars and dark matter), such that the dynamical friction force can be calculated \citep[see][]{BinneyTremaine2008} as 
\begin{equation}
    \textbf{F}_{\rm{DF}} = -4 \pi \rho_{sph} \left ( \frac{G M_{\rm{BH}}}{v_{\rm{BH}}} \right )^2 {\rm{log}} (\Lambda) \mathcal{F} \left ( \frac{v_{\rm{BH}}}{\sigma_v} \right ) \frac{\textbf{v}_{\rm{BH}}}{v_{\rm{BH}}}\,,
\end{equation}
where $\rho_{\rm{sph}}$ and $\sigma_v$ are the density and velocity dispersion of the surrounding dark matter and star particles, $\textbf{v}_{\rm{BH}}$ is the velocity of the black hole relative to the surrounding medium, $\Lambda$ is the Coulomb logarithm 
\begin{equation}
    \Lambda = \frac{b_{\rm{max}}}{(GM_{\rm{BH}})/v_{\rm{BH}}^2}
\end{equation}
with $b_{\rm{max}}=20kpc$, and $\mathcal{F}$ is
\begin{equation}
    \mathcal{F} = {\rm{erf}}(x)-\frac{2x}{\sqrt{x}} e^{-x^2}, x=\frac{v_{\rm{BH}}}{\sigma_v}
\end{equation}
from integrating the Maxwellian distribution. This dynamical friction implementation produces physically realistic motion for black holes due to small scale interactions with the nearby matter, and stabilizes the black hole once it reaches the galactic center, which provide more realistic information for the black holes leading up to mergers.  For a more detailed discussion of this implementation and the black hole orbital information it produces, see \citet{Ni2022, Chen2022Astrid}.

\section{Black hole / merger populations}
\label{sec:merger_populations}

\begin{figure}
    \centering
    \includegraphics[width=8.5cm]{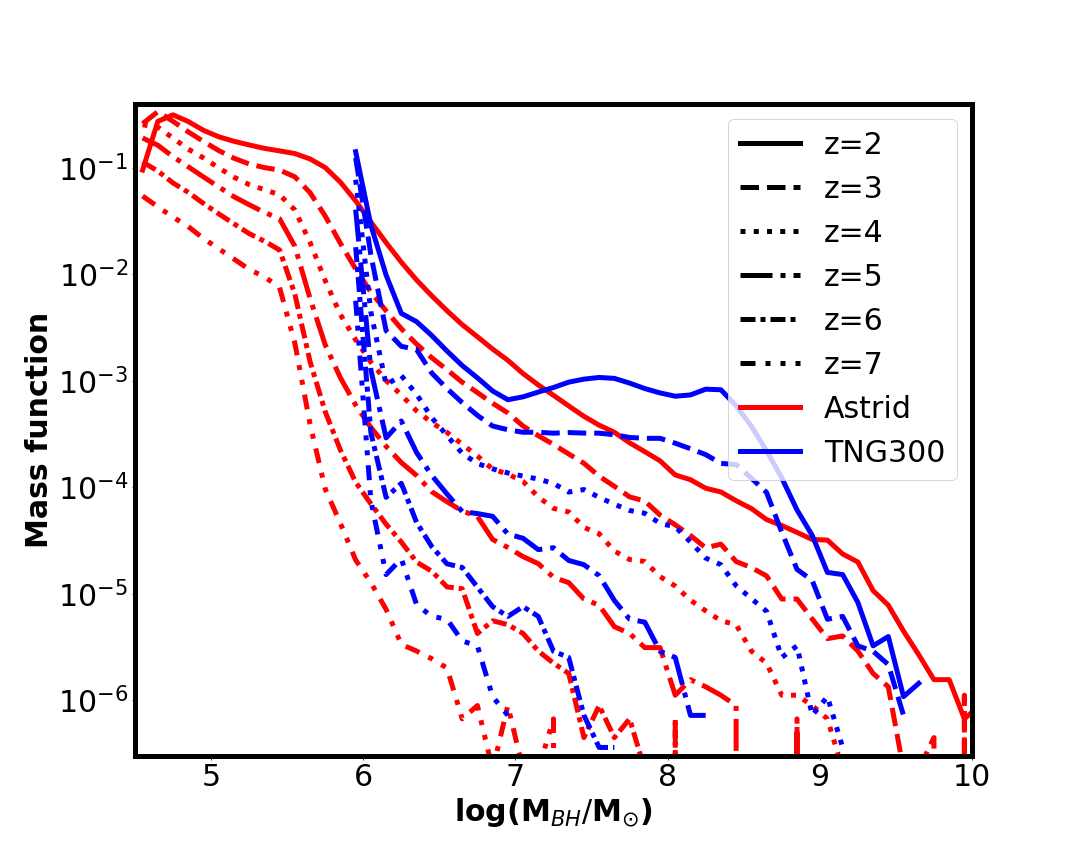}
    \caption{The black hole mass function for {\small{Astrid}} (red) and TNG300 (blue), at z=2-7.  For each simulation, the black hole seed model is clearly seen at the lowest masses, but above the seed mass the two simulations broadly agree for $z>4$, though TNG300 produces more high-mass black holes at late times.} 
    \label{fig:massfunction}
\end{figure}

Before investigating mergers in the {\small{Astrid}} simulation, we first consider the overall populations of black holes over cosmic time.  In Figure \ref{fig:massfunction} we plot the black hole mass function (BHMF) at $z=$2, 3, 4, 5, 6, and 7 (solid, dashed, dotted, and dot-dashed, respectively) from the {\small{Astrid}} simulation (red), with a comparison to the TNG300 mass function (blue) at the same redshifts.  We find the high end of the mass function ($M_{\rm{BH}} > 3 \times 10^6 M_\odot$) tends to follow an approximate power law with a slope of $\sim-1.05$ (for $M_{\rm{BH}} > 10^6 M_\odot$).  The slope is slightly steeper at earlier times, but the primary evolution is the increase in normalization (increasing by more than 2 dex from z=7 to z=2; see Table \ref{tab:bhmf_parameters} for best fitting power-law parameters at each redshift).  The TNG300 BHMF is broadly consistent with these {\small{Astrid}} results, except that TNG300 produces more high-mass black holes at later times ($\sim 3 \times$ more $\sim 10^8 \ h^{-1} \ M_\odot$ black holes at z=4).  As such, we can see that massive black holes are able to grow more efficiently in TNG than {\small{Astrid}} at late times (except for the most-massive end), which we can expect to have some impact on the merging populations as well.  However, we note that this will only affect the largest black holes at the latest times, and thus will not influence the majority of mergers when comparing the two simulations.

At lower masses, we see that the two simulations diverge significantly, as a result of the seeding model.  In the TNG300 simulation, we see a large spike at the lowest mass BHs ($\sim 10^6 M_\odot$).  This peak corresponds to the black hole seed mass used in the TNG simulations, and is a result of recently-seeded black holes tending to have very low accretion rates due to stellar feedback, especially at early times \citep{Weinberger2017, Weinberger2018}; hence a large number of black holes have not grown much beyond their seed masses.  In contrast, the {\small{Astrid}} BHMF gets slightly steeper at lower masses, but does not have any qualitative shift until $M_{\rm{BH}} \sim 10^{5.5} M_\odot$, below which we find a shallower power law.  We recall that the seeding model used in {\small{Astrid}} initializes black holes with a starting mass selected from a power-law distribution with slope of -1 (see Section \ref{sec:method}), which produces the behaviour we see here.  We note that the best fitting slope for this plateau is slightly shallower than the slope for the seed mass selection ($\sim -0.9$ rather than the $-1$ used when seeding), and gets gradually shallower with time (from $\sim -0.96$ at z=7 to $\sim -0.5$ at z=2; see also Table \ref{tab:bhmf_parameters}).  This is due to lower-mass seeds gradually growing into the higher mass range for seeding, and hence over time the number of black holes in the high-mass end of the seed range increases.  As such the plateau we see in the BHMF is a result of the seed model used combined with a small amount of growth, while for higher-mass black holes (e.g. $> 10^6 M_\odot$) we find a well-behaved BHMF which is consistent with the TNG BHMF.

\begin{figure}
    \centering
    \includegraphics[width=8.5cm]{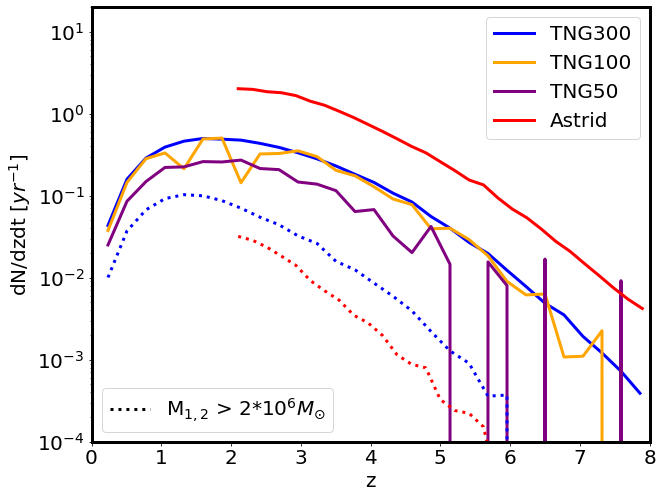}
    \caption{Merger rate signal as a function of redshift in {\small{Astrid}} (red), TNG300 (blue), TNG100 (yellow), and TNG50 (purple).  Since {\small{Astrid}} includes much lower mass black holes than the TNG simulations, the dashed red line shows the {\small{Astrid}} signal rate when limited to only mergers massive enough to be resolved in TNG300.  We see that all five simulations have qualitatively similar behaviour, though with different normalization: {\small{Astrid}} has significantly more mergers (due to the lower-mass black holes).  When considering only mergers between blackholes above $2 \times M_{\rm{TNG,seed}}$ (dotted lines), however, we find a \textit{lower} merger rate in Astrid. }
    \label{fig:merger_zdist}
\end{figure}

\begin{figure*}
    \centering
    \includegraphics[width=17cm]{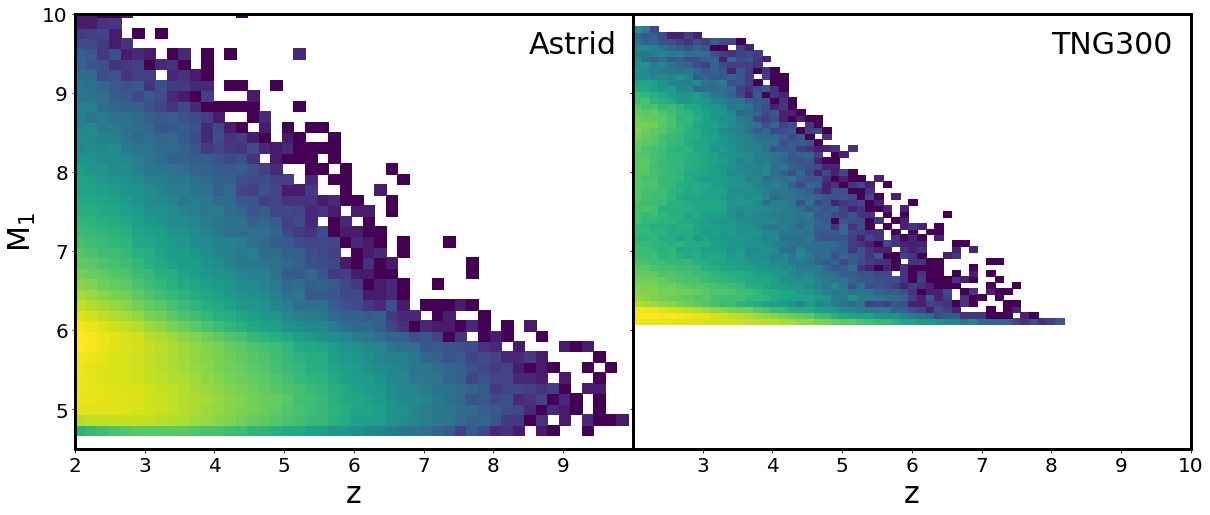}
    \caption{Distribution of primary merger mass ($M_1$) and redshift for {\small{Astrid}} (left) and TNG300 (right; restricted to $z>2$ to match {\small{Astrid}}).  The different seed models allows {\small{Astrid}} to probe to lower masses and higher redshifts, but produce comparable results above $M_{\rm{seed,TNG}}$.}
    \label{fig:M1_vs_z}
\end{figure*}

Overall, we find that {\small{Astrid}} is similar to TNG300 (and comparable simulations), though with two significant differences.  First, {\small{Astrid}} spans a larger range of black hole masses, extending to much smaller black holes resulting from the new seed model.  Additionally, {\small{Astrid}} has an improved black hole treatment, including dynamical friction, which more accurately models black hole motion and mergers. As such, we expect the merger rates to be more strongly affected, and a potential impact on high-mass black holes (after longer time to grow), hence the discrepancy between Astrid and TNG300 in Figure \ref{fig:massfunction} increases for late times and high masses.

Next we consider the mergers produced by this black hole population.  In Figure \ref{fig:merger_zdist} we plot the rate at which GW signals will reach the Earth from SMBH mergers, obtained by integrating the number of mergers in the simulation over redshifts, incorporating the cosmic volume at the given redshift:
\begin{equation}
    \frac{{\rm d}N}{{\rm d}z\,{\rm d}t} = \frac{1}{z_2-z_1} \int_{z_1}^{z_2} \frac{{\rm d}^2 n(z)}{{\rm d}z\,{\rm d}V_c} \frac{{\rm d}z}{{\rm d}t} \frac{{\rm d}V_c}{{\rm d}z} \frac{{\rm d}z}{1+z}\,
\end{equation}

We see that the TNG simulations (TNG300 - blue; TNG100 - yellow; TNG50 - purple) have broadly comparable behaviour: very rare mergers at high z, increasing with time to a peak at z$\sim$2, followed by a decrease in expected merger rates.  The high-redshift behaviour in {\small{Astrid}} (red) is broadly similar, in that it starts with rare mergers at high z, and increases with time.  However, we find an overall rate significantly higher in {\small{Astrid}} than any of the TNG simulations.  Rather than a fundamental difference in mergers, however, this increased rate in {\small{Astrid}} is a result of the seed model initializing lower mass black holes in lower mass galaxies when compared to TNG. As shown in Figure \ref{fig:massfunction}, at high masses the simulations are consistent, but {\small{Astrid}} includes many low-mass BHs below TNG's seed mass, and hence will include a significant number of mergers which go unresolved in TNG. 

\begin{table}
    \centering
    \begin{tabular}{c|c|c|c|c}
     \hline
       &  \multicolumn{2}{c}{\underline{High-mass}} & \multicolumn{2}{c}{\underline{Low-mass}} \\
     z &  $a$ & $b$ & $c$ & $d$ \\
     \hline
     2 & -2.7 & -1.0 & -1.6 & -0.49 \\
     3 & -3.3 & -1.0 & -2.0 & -0.60 \\
     4 & -3.9 & -1.1 & -2.5 & -0.79 \\
     5 & -4.6 & -1.0 & -2.9 & -0.87 \\
     6 & -5.4 & -1.1 & -3.3  & -0.93\\
     7 & -6.3 & -1.1 & -3.6 & -0.96 \\
     \hline
     
    \end{tabular}
     \caption{Best fitting parameters for black hole mass function (Figure \ref{fig:massfunction}), fitting $M_{\rm{BH}} > 10^6 M_\odot$ to $10^a (M/10^7 M_\odot)^b$, and $M_{\rm{BH}} < 10^{5.5} M_\odot$ (seed-mass range) to $10^c (M/10^7 M_\odot)^d$.}
\label{tab:bhmf_parameters}

\end{table}

To test this, the dashed lines in Figure \ref{fig:merger_zdist} show the rates from the {\small{Astrid}} and TNG300 simulations after imposing a cut of $2\times M_{\rm{seed,TNG}}$ (i.e. only mergers with $M_1 > M_{\rm{seed,TNG}}$ and $M_2 > M_{\rm{seed,TNG}}$, where $M_1$ ($M_2$) is the larger (smaller) mass involved in a merger, and $M_{\rm{seed,TNG}}$ is the black hole seed mass for the TNG simulations). A cut of $M > 2 \times M_{\rm{seed,TNG}}$ removes black holes which are not yet seeded by TNG, and also ignores the strong peak in the TNG BH population caused by the seed criteria, and thus we are comparing equivalent black hole populations.  Here we see that the majority of mergers involve low mass BHs (where at least one BH has $M_{\rm{BH}} < 2\times M_{\rm{seed,TNG}}$, hence dashed line well below the solid lines), and when limited only to the equivalent merger populations, {\small{Astrid}} actually has a \textit{lower} merger rate than the TNG simulations by a factor of $\sim 3$.  

There are several major differences between the simulations which have the potential to influence the merger rates.  Once factor is that {\small{Astrid}} seeds black holes into smaller halos, and the dependence of galaxy merger rates on galaxy mass will influence the rate at which BHs seeded in those galaxies end up merging.  However, this will be limited primarily to low mass galaxies/BHs, and so would not explain the lower rate in {\small{Astrid}} among high-mass mergers.  Another factor is the faster growth in TNG, which produces a larger population of high-mass black holes when compared to {\small{Astrid}} (see Figure \ref{fig:massfunction}); at $z=4$ TNG has $\sim 50\%$ more black holes above $2\times M_{\rm{seed,TNG}}$.  Hence we should expect a higher merger rate in TNG than {\small{Astrid}}, based solely on the population of black holes available to merge. 
Finally, there is the dynamical friction model: {\small{Astrid}} models black hole motion using a dynamical friction prescription \citep{Chen2022} rather than a recentering scheme. As such, when a pair of galaxies merge together, satellite BHs (those found in the smaller of the two merging galaxies) can take longer to reach the galaxy center where they are then able to merge with the central BH, and the full dynamical friction model allows for flyby interactions rather than assuming an immediate merger when two black holes are sufficiently close together.  Thus we expect black holes in simulations which incorporate dynamical friction to generally merge more slowly than those in simulations which use a recentering scheme. They should therefore have a lower overall merger rate (when controlling for resolved masses), precisely as we see when comparing {\small{Astrid}} to TNG300 (dotted lines in Figure \ref{fig:merger_zdist}).

To more directly consider merging masses, we plot the mass-redshift distribution of mergers in Figure \ref{fig:M1_vs_z} for both {\small{Astrid}} (left) and TNG300 (right).  Consistent with Figures \ref{fig:massfunction} \& \ref{fig:merger_zdist}, we see that the lower seed mass in {\small{Astrid}} provides $M_1$ more than an order of magnitude below what the TNG simulations resolve.  Similarly, the lower halo mass threshold for seeding means that {\small{Astrid}} models black hole mergers out to earlier cosmic times, as a result of seeding black holes into smaller galaxies.  Above the seed mass scales, however, we see that both {\small{Astrid}} and TNG300 produce remarkably similar distributions, with $M_1 > M_{\rm{seed,TNG}}$ mergers first occurring at $z \sim 7$, and reaching $M_1 \sim 10^{10} M_\odot$ at z=2 (though {\small{Astrid}}'s larger volume means there are a few unusually massive mergers at slightly earlier times).  This is consistent with Figure \ref{fig:massfunction}, in that black holes well above the seed mass tend to grow at comparable rates in both {\small{Astrid}} and the TNG simulations at high-z, and thus the mass scale involved in the mergers tends to be similar.  
Combined with Figure \ref{fig:massfunction}, we see that the smaller seeds in {\small{Astrid}} (and the power-law seed mass distribution) are capable of growing to the masses necessary to match high-redshift black hole observations and produce black hole populations fully consistent with current constraints. A complete analysis of typical black hole growth behaviours is beyond the scope of this paper, but will be discussed in an upcoming work.  

\begin{figure}
    \centering
    \includegraphics[width=8.5cm]{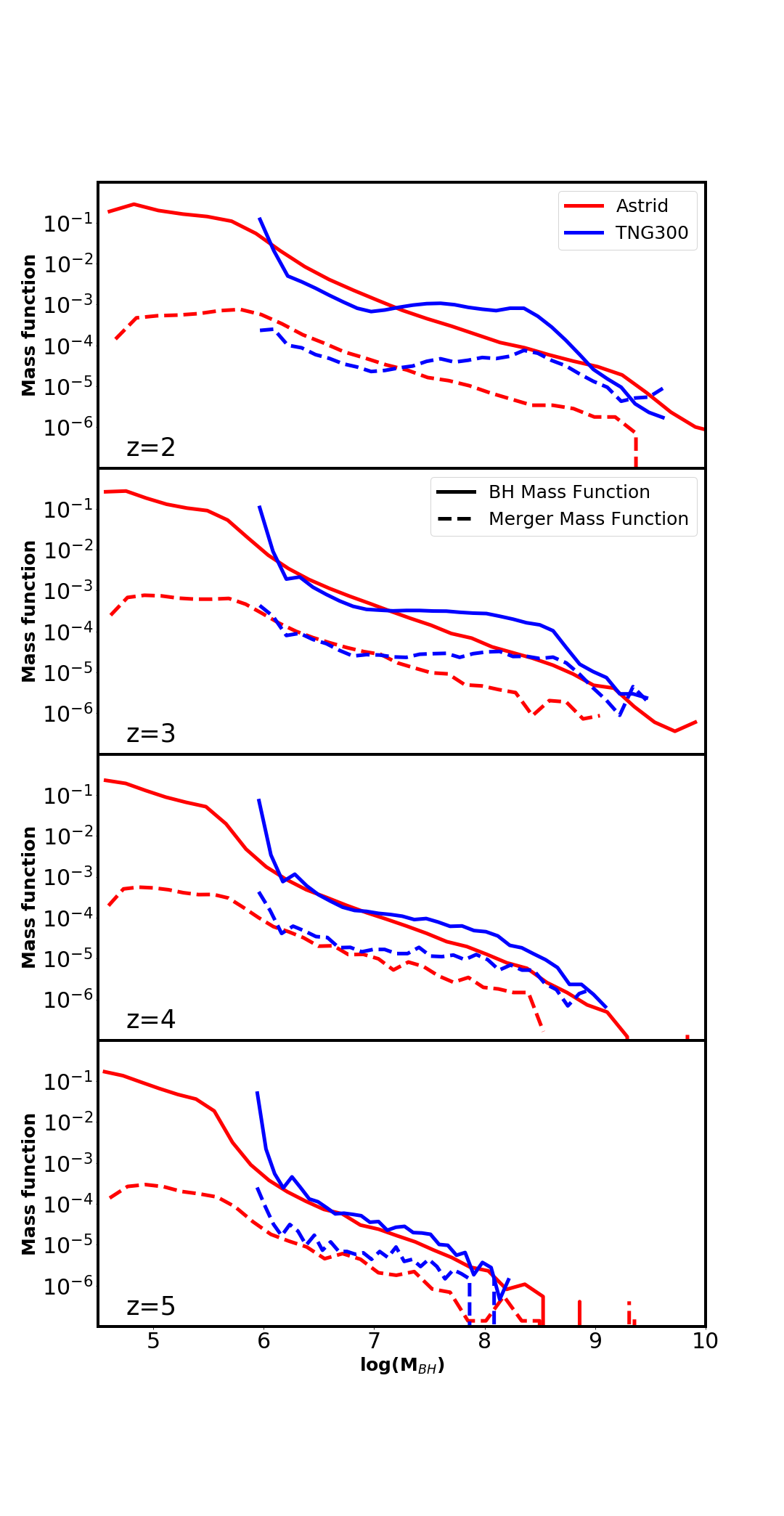}
    \caption{BHMF (solid) for {\small{Astrid}} (red) and TNG300 (blue) at z=2 (top) to z=5 (bottom), compared to the merger mass function based on the previous 75 Myr (dashed; see text for details).  In both simulations the merger mass functions are similar, though the {\small{Astrid}} merger mass function is smaller relative to the overall BHMF, suggesting a lower merger rate. }
    \label{fig:massfunction_withmergers}
\end{figure}

To compare merger rates to black hole populations, in Figure \ref{fig:massfunction_withmergers} we plot the black hole mass function (solid lines), and compare to the merger mass function (dashed line) at $z=2,3,4,5$.  We define the merger mass function to be the mass function of $M_1$ (the mass of the more massive BH involved in a merger) for all mergers which take place in the previous 75 Myr.  Here we see that the merger mass function in each simulation is comparable to (though slightly lower than) the corresponding BHMF.  In the case of TNG300, we see that above the seed mass both the BHMF and the merger mass function follow a rough power law, and both functions also show a peak at the seed mass (though that spike is smaller in the merger mass function).  Similarly, in {\small{Astrid}} we see a rough power law above $\sim 10^6 M_\odot$, and a plateau below $\sim 10^{5.5.} M_\odot$ which corresponds to the power-law seeding model.  Similar to TNG, we see that at the seed-mass scale the merger mass function is significantly below the BHMF: although there is a large population of recently seeded black holes, they are less likely to undergo a merger (as they necessarily need time for their host halos to merge, during which they are able to grow past their seed mass).

\begin{figure}
    \centering
    \includegraphics[width=8.5cm]{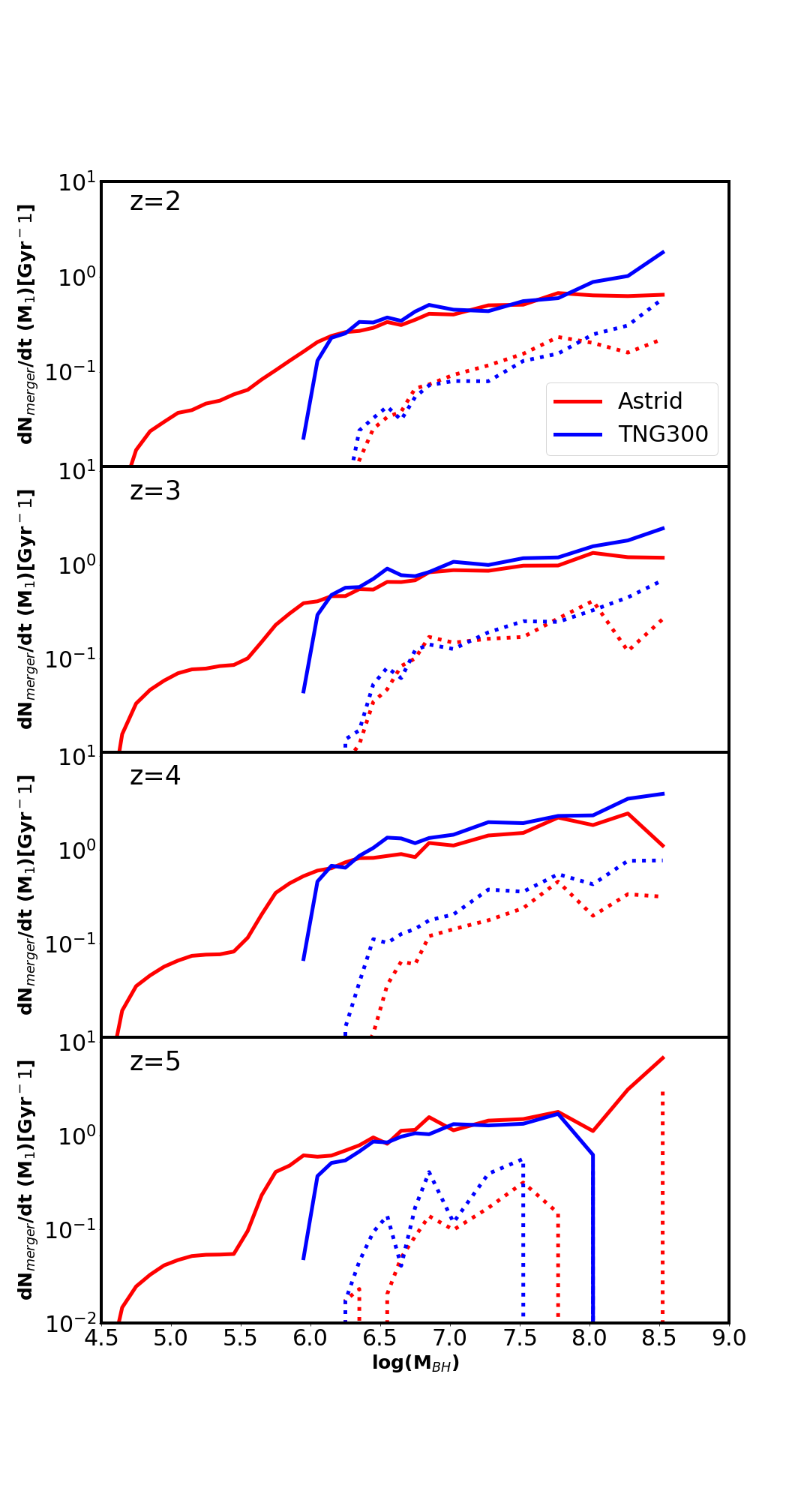}
    \caption{The typical merger rate for a BH with a given mass in the {\small{Astrid}} (red) and TNG300 (blue) simulations, as well as the merger rates limited to $M_{1,2} > 2*M_{\rm{seed,TNG}}$ (dotted lines).   Both simulations have comparable total merger rates, but the majority of mergers in {\small{Astrid}} involve a low-mass secondary BH; when restricted to the same secondary mass range, {\small{Astrid}} has a merger rate $\sim 1$ dex lower than TNG300. [NOTE: This is not the overall rate of mergers in the simulation/universe; rather it is the rate at which an individual BH of a given mass would be expected to undergo mergers; i.e. lower panel would suggest a z=5 BH with M$\sim 10^{6.5} M_\odot$ would be expected to undergo about 1 merger per Gyr.  That's true for both TNG and {\small{Astrid}}, but in {\small{Astrid}} that includes mergers with lower-mass BHs (below the seed mass of TNG); the dashed line only includes mergers where both BHs are above the seed mass of TNG, so it's closer to a 1-to-1 comparison.]}
    \label{fig:merger_timescale}
\end{figure}

\begin{figure*}
    \centering
    \includegraphics[width=17cm]{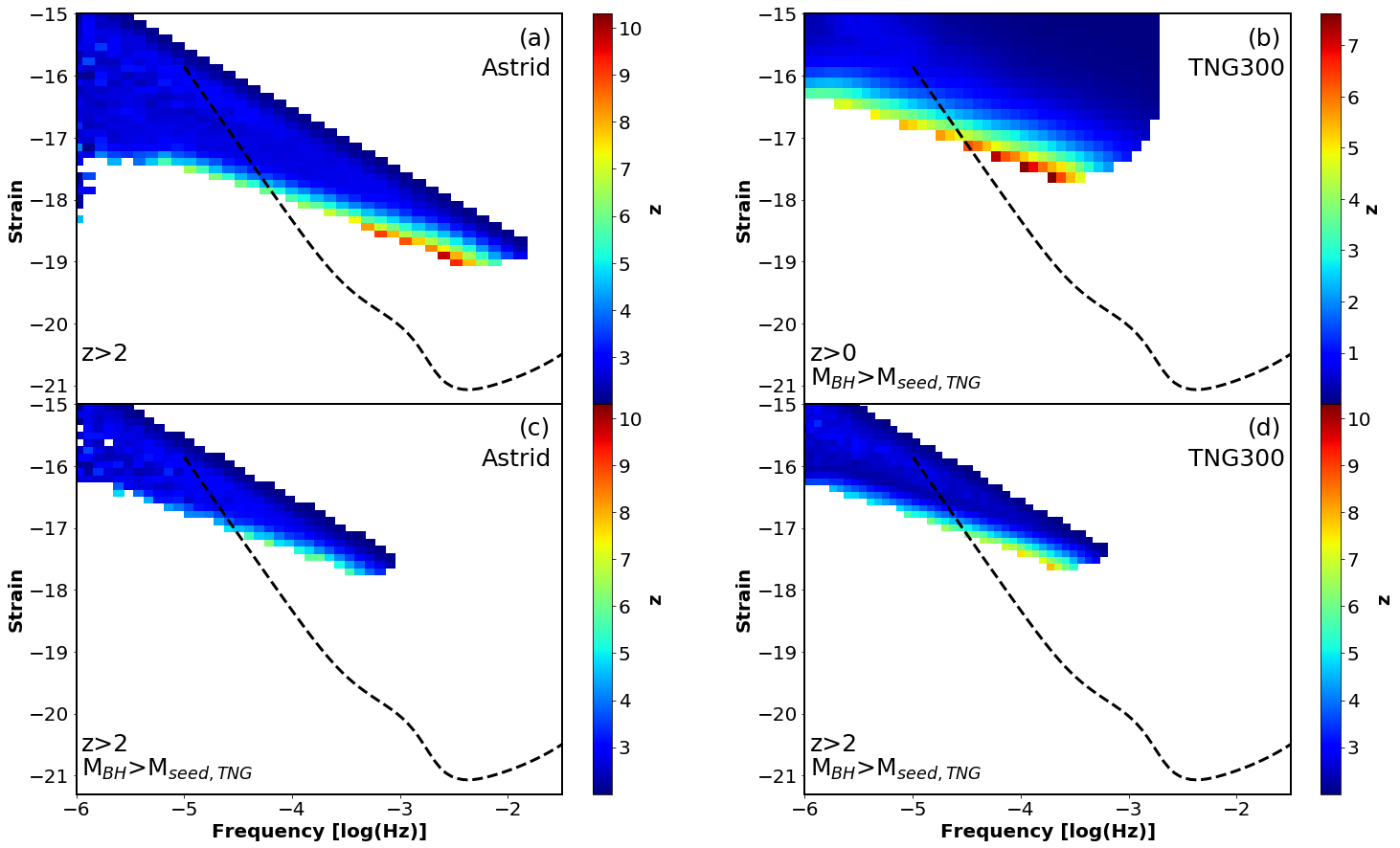}
    \caption{Frequency-strain signals for GWs emitted by BH mergers over cosmic time (binned data) against the LISA sensitivity curve (dashed black line).  Each bin is colour-coded by the mean redshift of the source merger which emitted the given GWs.  \textit{(a):} All signals from the {\small{Astrid}} simulation (limited to $z>2$). \textit{(b):} All signals from the TNG300 simulation, which is complete to z=0 (hence note the different colourscale range), but limited to higher masses. \textit{(c):} Signals from the {\small{Astrid}} simulation, but limited to the same mass scales as TNG300.  \textit{(d):} Signals from the TNG300 simulation, but limited to the same redshift range ($z>2$) as the current state of the {\small{Astrid}} simulation.} 
    \label{fig:freq_strain}
\end{figure*}

Although the ratio between BHMF and merger mass function is qualitatively similar between {\small{Astrid}} and TNG, suggesting a comparable merger rate, when selecting BHs by mass.  In Figure \ref{fig:merger_timescale}, we have divided the merger mass function by the BHMF to obtain a characteristic merger rate as a function of $M_{\rm{BH}}$ (i.e. the typical number of mergers that a BH with a given mass would undergo in 1 Gyr).  As discussed in reference to Figure \ref{fig:massfunction_withmergers}, we expect the merger rate for BHs near the seed mass to generally be much lower than the merger rate for higher mass black holes: we see this explicitly in Figure \ref{fig:merger_timescale}, as both {\small{Astrid}} (red) and TNG300 (blue) show a significant dropoff near their respective seed masses.  Above the seed masses, however, both simulations show an approximate power law relating merger rate to $M_{\rm{BH}}$, such that high-mass black holes have a slightly higher merger rate than lower-mass black holes.
Furthermore, we see that both {\small{Astrid}} and TNG300 have comparable merger rates; however, we note that this is somewhat coincidental, based on near seed-mass mergers.  As seen in Figures \ref{fig:massfunction}-\ref{fig:massfunction_withmergers}, {\small{Astrid}} contains a large number of low-mass black holes which are unresolved in TNG, but which contribute to the merger rates show in Figure \ref{fig:merger_timescale}.  We show this explicitly with the dashed line, where we only consider mergers in which $M_2 > 2\times M_{\rm{seed,TNG}}$, in which case we have roughly an order of magnitude fewer mergers.  Nonetheless, after removing the low-mass (and thus less well-resolved) mergers, we find fairly comparable merger rates except at the highest masses, where {\small{Astrid}} predicts a slightly longer time between mergers.  

\section{Gravitational Wave Signals}
\label{sec:GWs}

In addition to the underlying black hole merger populations, we consider the gravitational waves emitted by calculating both the frequency and strain of the GW for each merger.  We use the characteristic strain, $h_s$, to model the binary signal which accounts for the time the binary spends in each frequency bin \citep[][]{Finn2000}. 
The characteristic strain is given by \citep[e.g.][]{Moore2015}:
\begin{equation}
\label{eq:hc}
    h_s(f) = 4f^2 |\tilde{h}(f)|^2
\end{equation}
 where $\tilde{h}(f)$ represents the Fourier transform of a time domain signal.
 To generate the waveforms, we use the phenomenological waveform PhenomD \citep[][]{Husa2016,Khan2016} implemented within the \texttt{gwsnrcalc} Python package \citep[][]{Katz2019}. 
 The input parameters are the binary masses, merging redshift, and the dimensionless spins of the binary. 
 For the SMBH masses, we do not account for mass growth after the numerical merger. 
 However, we note that the SMBH can potentially gain a significant fraction of its mass during the $>1\,{\rm Gyr}$ of time in the dynamical friction \citep[e.g.][]{Banks2022} or loss-cone scattering phase. 
 The dimensionless spin $a$ characterizes the alignment of the spin angular momentum with the orbital angular momentum, and the value of $a$ ranges from $-1$ to $1$. 
 However, we do not have any information on the spin of the SMBHs in our simulation. 
 Therefore, following the argument in \citet{Katz2020}, we assume a constant dimensionless spin of $a_1=a_2=0.8$ for all binaries \citep[e.g.][]{Miller2007,Reynolds2013}.

In Figure \ref{fig:freq_strain} we plot the range of frequencies and strains for GW signals emitted by mergers in {\small{Astrid}} (panel a), with each frequency-strain bin colour-coded by the mean redshift of the emitting mergers.  We see that the majority of GW signals come from the lowest-redshift mergers (though note that the lowest redshift here means $z \sim 2$, the latest time probed in this analysis), since the merger rate increases with time (at least at high-z; see Figure \ref{fig:merger_zdist}).  We note that high-z does dominate the lowest-strain GW signals, which is not only a direct result the merger being more distant (and hence a weaker signal), but also because high-z mergers tend to involve swallowing a lower-$M_2$ BH (which necessarily produces a smaller strain compared to swallowing a higher-mass secondary BH).  We also see an upper bound on the GW signals, however we note that is a result of the limited redshift window reached so far by {\small{Astrid}} (as the simulation continues to lower-z, the upper right area will continue to be filled in, which we see in panels $b$ and $d$ for TNG300 mergers).

In Figure \ref{fig:freq_strain}b, we show the frequency-strain signals for the TNG300 mergers, again colour-coded by the merger redshift (though note the range for the colour bar is different, since TNG is complete to z=0).  As in Figure \ref{fig:freq_strain}a, we see that most signals come from the lowest redshifts (although here that means near $z=0$), and again high-z signals dominate the lowest-strains.   However, the much larger redshift range makes comparing to these panels problematic; instead, in Figure \ref{fig:freq_strain}d we plot the GW signals from TNG300, but restricted to the same redshift range as {\small{Astrid}} ($z > 2$).  Here we see a result which is more similar to the {\small{Astrid}} data (Figure \ref{fig:freq_strain}a), but which spans a much more limited range of frequency and strain.   
The z$\sim$2 limit imposes a strict upper limit on the strain, and we see that both {\small{Astrid}} (Figure \ref{fig:freq_strain}a) and TNG300 (Figure \ref{fig:freq_strain}d) have similar distributions when limited to the same redshift range, with the exception that {\small{Astrid}} extends higher frequencies and lower strain, which we expect since {\small{Astrid}} includes black holes below the resolved mass in TNG300.   
Overall, we see that the GW signals tend to be dominated by the lowest redshift mergers, with an exception for the highest frequency/lowest strain signals, which tend to be at higher redshifts.

Unlike Figure \ref{fig:freq_strain}a, which spans the majority of the LISA frequency range, the frequency-strain distribution for TNG300 (Figure \ref{fig:freq_strain}b \& d) is limited only to LISAs low-frequency regime.  This is a result of {\small{Astrid}} resolving much lower black hole masses, whose mergers produce higher frequency signals. We see this explicitly in Figure \ref{fig:freq_strain}c, which shows the GW signals in {\small{Astrid}}, but limited to the same mass range as TNG300 (i.e. $M_2 > M_{\rm{seed,TNG}}$), such that Figure \ref{fig:freq_strain}c \& d show the results from {\small{Astrid}} and TNG300 for the same mass and redshift ranges.  Comparing these panels, we see that {\small{Astrid}} has fewer mergers (at fixed mass/redshift ranges) than TNG300, consistent with Figures \ref{fig:merger_zdist} \& \ref{fig:merger_timescale} which also show fewer mergers.  Since the mass scale is well above the seed mass, this is not a result of the seeding prescriptions.  Rather it is a result of the dynamical friction modeling which delays black hole mergers to later times relative to a simpler repositioning scheme.  Overall, we see that the largest difference between the two simulations is that {\small{Astrid}} produces more high-frequency and low-strain signals, which correspond to low-mass mergers.

\begin{figure}
    \centering
    \includegraphics[width=8.75cm]{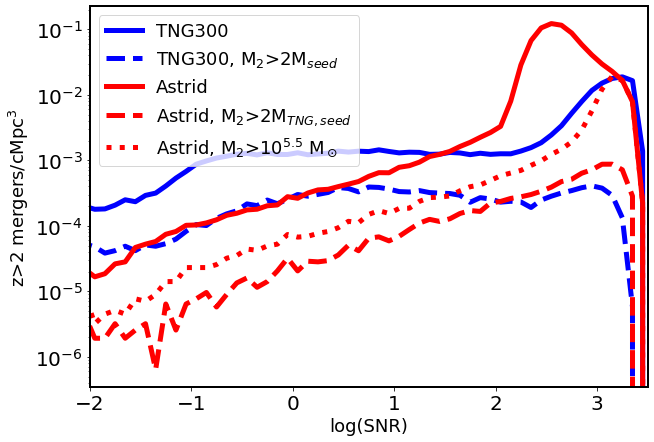}
    \caption{Distribution of signal-to-noise ratios (SNR) for $z>2$ mergers from {\small{Astrid}} and TNG300, normalized by simulation volume (solid lines).  Both simulations show qualitatively similar behaviour, with the distribution peaking at high SNR. Because the simulations use different seed masses, we also compare the SNR distribution for mergers in which both merging black holes are above $2 \times$ the TNG seed mass (dashed lines), as well as the distribution of {\small{Astrid}} signals for mergers in which both merging black holes are above $10^{5.5} \ h^{-1} \ M_\odot$ (the high-end of {\small{Astrid}}'s seed mass range).  }
    \label{fig:mergerhist_SNR}
\end{figure}

\begin{figure*}
    \centering
    \includegraphics[width=17cm]{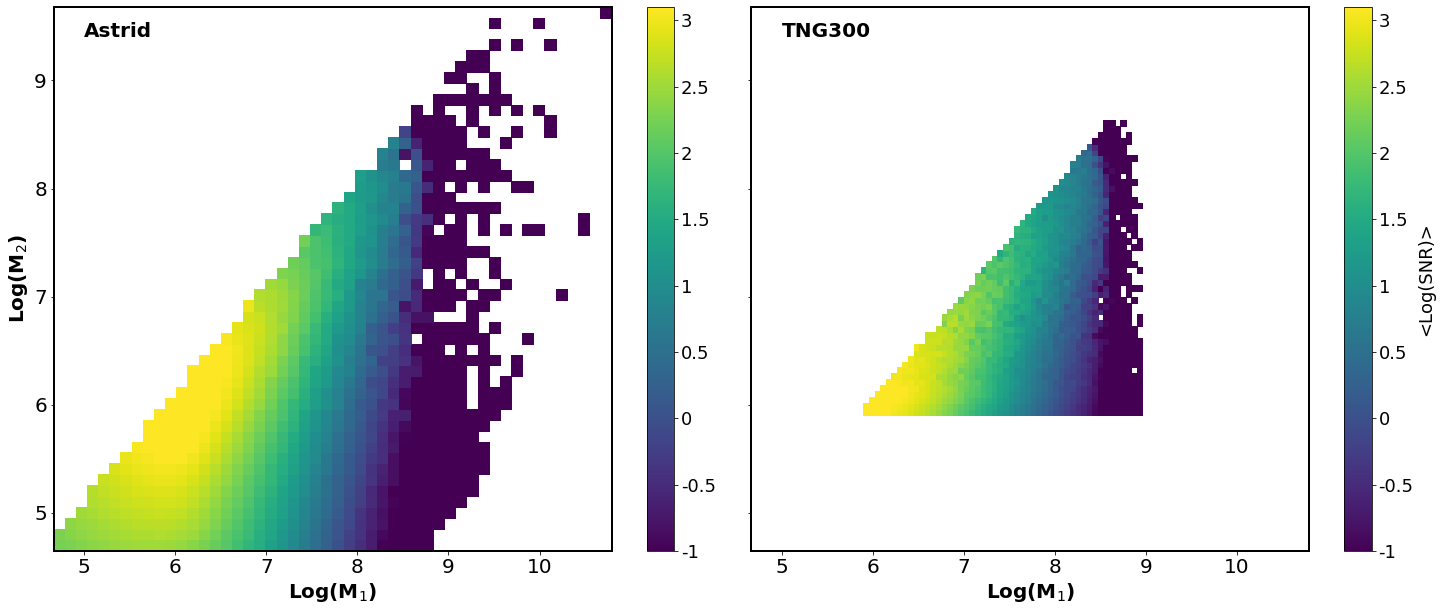}
    \caption{Distribution of merging black hole masses for {\small{Astrid}} (left) and TNG300 (right), color coded by the mean log(SNR) for the mergers.  We find that the strongest SNRs in {\small{Astrid}} are found for $M_1, M_2 \sim 10^6 M_\odot$; this is consistent with TNG300, although the higher seed mass in the TNG simulations means that the peak signals roughly correspond to the seed masses of the simulation.  In both simulations, we also note that the SNR is more strongly correlated to $M_1$ than $M_2$. }
    \label{fig:M1M2_SNR}
\end{figure*}

\subsection{SNR}
\label{sec:SNR}

Having calculated the GW frequency and strain emitted by merging black holes, we also consider the strength of the GW signal received by LISA.  We estimate the Signal-to-Noise (SNR) by integrating the ratio of the signal to the noise in the frequency domain. The sky, orientation, and polarization averaged SNR are given by :
\begin{equation}
\label{eq:snr}
\langle {\rm SNR} \rangle^2 = \frac{16}{5}\int_{f_{\rm start}}^{f_{\rm end}} \frac{h_s^2}{h_N^2} f^{-1} df,
\end{equation}
where $f_{\rm start} = f(t_{\rm start})$ and $f_{\rm end} = f(t_{\rm end})$, with $t_{\rm start}$ and $t_{\rm end}$ representing the starting and ending time when the signal is observed. Note that here we do not account for the eccentricity of the binaries, and assume circular orbits at the time of merger.

\begin{figure*}
    \centering
    \includegraphics[width=17cm]{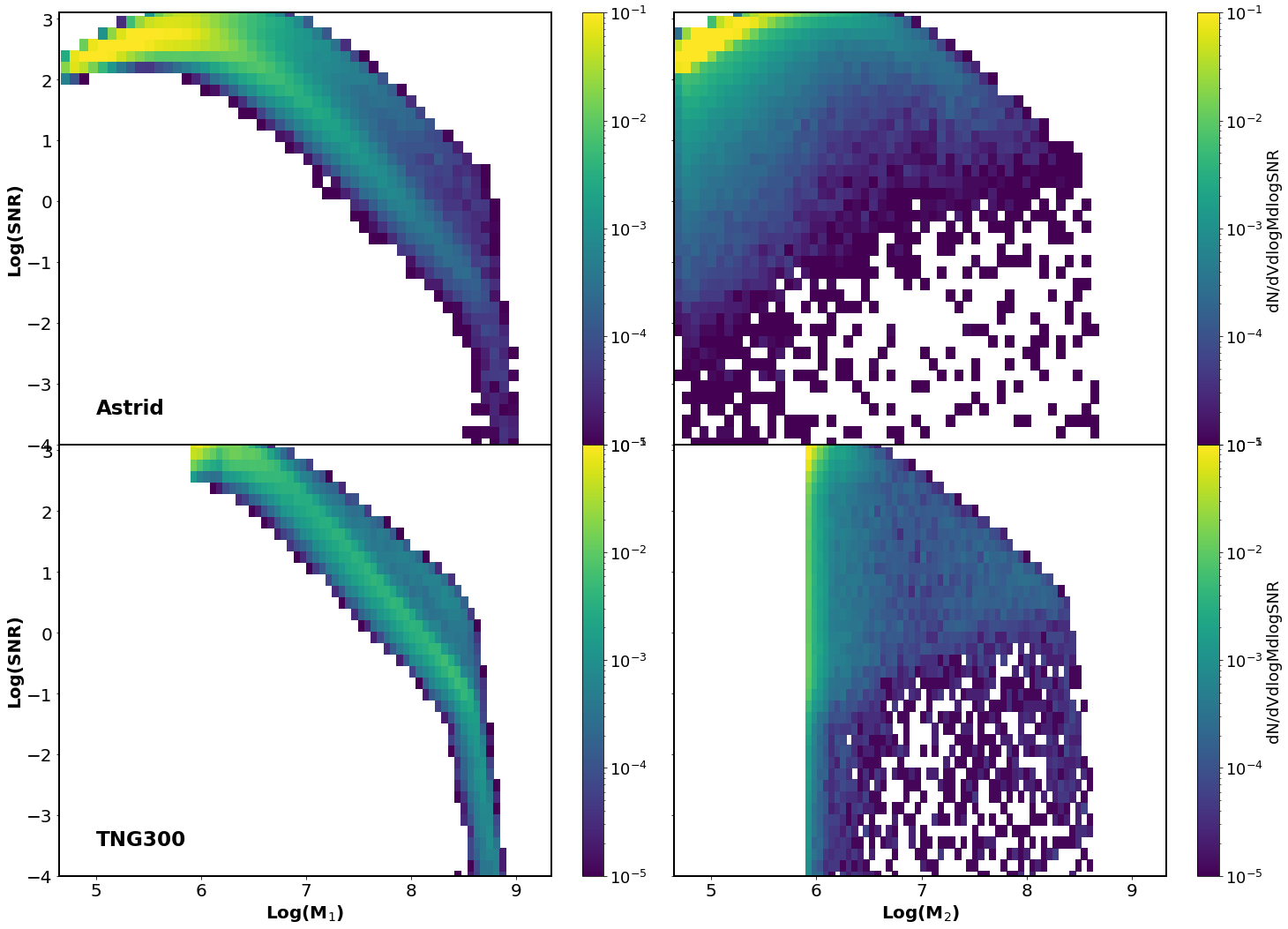}
    \caption{Distribution of SNR vs. mass (left: $M_1$; right: $M_2$) for $z>3$ mergers in {\small{Astrid}} (top) and TNG300 (bottom), color-coded by density of mergers in the simulation volume. Consistent with Figure \ref{fig:M1M2_SNR}, we see that SNR is much more strongly correlated wtih $M_1$ than $M_2$, and we again find that {\small{Astrid}} and TNG300 are broadly consistent above the TNG seed mass.}
    \label{fig:mass_SNR_4panel}
\end{figure*}

For the current configuration, we assume that the LISA observation lasts for 4 years. We further assume a most optimistic SNR for all mergers by taking $t_{\rm end}=t_{\rm peak}$ and $t_{\rm start}=t_{\rm peak}-4{\rm yrs}$. Under this assumption, we are always integrating the part of the waveform where the strain is maximized. However, as was discussed in \citet{Salcido2016} and \citet{Katz2020}, the actual SNR may be smaller if there is an offset between the LISA observation window and the merger time of the binary.

In Figure \ref{fig:mergerhist_SNR}, we plot the distribution of SNRs from {\small{Astrid}} (solid red) and TNG300 (solid blue) for $z>2$ mergers, normalized by the simulation volume.  In both simulations we find that the SNR peaks at very high values ($10^2-10^3$), with {\small{Astrid}} producing significantly more high-SNR mergers, while TNG300 has a flatter tail extending toward low-SNR.  Since the majority of mergers tend to occur between low-$M_{\rm{BH}}$ black holes (e.g., see Figures \ref{fig:M1_vs_z}-\ref{fig:massfunction_withmergers}), these differences are expected due to the different seed criteria between the two simulations.  For a more direct comparison, the dashed lines show the SNR distribution limited to mergers in which both merging black holes have $M_{\rm{BH}} > 2 \times M_{\rm{TNG,seed}}$ (i.e. only including BHs more than double the TNG seed mass, to avoid issues related to black hole seeding).  Here we see that the two simulations have a consistent peak at SNR just below $10^3$, though we find that TNG300 has a nearly flat distribution for SNR$>$1, whereas {\small{Astrid}} predicts a moderate slope, producing fewer low-SNR mergers.  This appears to be due to TNG300 producing more high-mass black holes at $z \le 4$ (see Figure \ref{fig:massfunction_withmergers}) as a result of more efficient black hole growth.  A detailed comparison of black hole growth efficiency between these simulations is beyond the scope of this paper, and is left for a future work.  However, we note that the difference only affects the low-SNR end, and will thus have a comparatively minor impact on the LISA detection rate.

On the other hand, the high-SNR end is clearly significantly impacted by the seeding criteria.  Although {\small{Astrid}} and TNG agree when limited to merging masses about $2 \times M_{\rm{TNG,seed}}$, the majority of mergers are removed when applying such a high mass cut.  One of the major advantages of the {\small{Astrid}} simulation is that the lower seed masses allow us to effectively probe lower mass mergers.  We show this with the dotted line in Figure \ref{fig:mergerhist_SNR}, where we consider any mergers between BHs above the {\small{Astrid}} seed mass distribution (i.e. $M_{\rm{BH}} > 10^{5.5} \ h^{-1} \ M_\odot$, hence above the seed-mass dominated regime, as seen in Figure \ref{fig:massfunction}), which provides us with significantly more mergers.  We note that including these lower-mass black holes produces a peak at the high-SNR end which is missed when imposing a higher mass cut (or higher seed masses). Furthermore, we find roughly triple the number of SNR$< 10^{2.5}$ for $M_2 > 10^{5.5} \ h^{-1} \ M_\odot$ than for $M_2 > 2 \times M_{\rm{seed,TNG}}$, emphasizing the importance of low-mass black holes.  Furthermore, we note that the slope of the SNR$< 10^{2.5}$ distribution remains roughly the same when using the lower mass cut.  Overall, this suggests that the highest-SNR mergers are primarily between two low-mass BHs which are completely missed with a higher mass cut (hence the strong peak which the dashed line completely misses), while the lower-SNR mergers frequently involve at least one higher-mass black hole, so by including a smaller mass cut we increase the number of mergers at all SNR-ranges but without affecting the slope.

We investigate this in more detail in Figure \ref{fig:M1M2_SNR}, which shows the mass distribution of merging black holes ($M_1$ vs. $M_2$), color coded by $<\rm{log}(SNR)>$ (i.e. the mean logarithmic SNR for mergers in the given mass bin), for $z>3$ mergers in {\small{Astrid}} (left) and TNG300 (right).  In the mass scales resolved in both simulations we find comparable SNR results (though we again see more high-mass mergers in TNG as a result of the more efficient growth).  As seen in Figure \ref{fig:mergerhist_SNR}, the larger mass range produced by {\small{Astrid}}'s smaller seed mass results in significantly more mergers, especially for high SNRs.  In particular, the {\small{Astrid}} simulation shows the SNR peaks at $M_1, M_2 \sim 10^6 M_\odot$, with the lower masses having weaker (though still strong) signals (they are below LISA's peak sensitivity, but still well within the detectable range).  

Figure \ref{fig:M1M2_SNR} also shows that SNR depends primarily on the larger mass ($M_1$), while the secondary black hole ($M_2$) has a relatively weak impact on SNR.  We show this explicitly in Figure \ref{fig:mass_SNR_4panel}, which plots the correlation between SNR and the merging black hole masses ($M_1$ - left; $M_2$ - right) for {\small{Astrid}} (top) and TNG300 (bottom), color coded by the number density of mergers.  The majority of mergers occur between low mass black holes (as seen, e.g., in Figure \ref{fig:massfunction_withmergers}), and we again see that the highest SNRs occur when both $M_1$ and $M_2$ are $\sim 10^6 M_\odot$.  We also see that SNR correlates quite strongly with $M_1$, spanning $\sim$1 dex at $M_1 \sim 10^6 M_\odot$ up to $\sim$2 dex at higher masses.  In contrast, a low $M_2$ spans a much higher range of SNRs, corresponding to the wide range of possible $M_1$ (i.e. mass of the swallowing black hole), though most low-$M_2$ mergers still have high-SNR (since the majority of mergers will still have a low $M_1$).   The relatively weak dependence on $M_2$ means that the growth of the secondary BH does not play a significant role in the expected SNR for a given merger.  However, it is nonetheless very important to include lower mass black holes to probe the full range of mergers which we expect LISA to be able to detect: mergers between $\sim 10^6 M_\odot$ black holes are both extremely common and produce the strongest gravitational wave signals from LISA, so limiting simulations to higher masses will necessarily miss the majority of mergers and their associated GW signals.  On the other hand, we note that the LISA SNR peaks at both $M_1 \sim 10^6 M_\odot$ and $M_2 \sim 10^6 M_\odot$.  So although incorporating smaller seed masses in simulations will provide more accurate estimates for the overall merger rates (which are dominated by low-mass mergers), below $\sim 10^5 M_\odot$ will primarily consist of lower-SNR signals.

\section{Conclusions}
\label{sec:conclusions}

In this work, we have investigated the black hole population in the {\small{Astrid}} simulation, which seeds black holes at masses as low as $10^{4.5} \ h^{-1} \ M_\odot$, well below what is frequently used in comparable cosmological simulations.  We have particularly focused on the mergers between black holes, looking at the overall merger rates, the merging masses, and the expected gravitational wave signals they produce, noting that {\small{Astrid}} directly incorporates a dynamical friction model which produces more realistic merger behaviour.  Our main results are as follows:

\begin{itemize}
    \item The {\small{Astrid}} simulation produces a comparable population of high-redshift black holes when compared to prior simulations (esp. TNG300), but with an alternate seed model which extends to much lower mass black holes.  Above the masses resolved by both simulations, we see that {\small{Astrid}} and TNG have comparable mass functions (Figure \ref{fig:massfunction}), and also typical merging masses as a function of redshift (Figure \ref{fig:M1_vs_z}).
    
    \item The overall merger rate is much higher in {\small{Astrid}} than in earlier TNG simulations, primarily due to mergers involving low-mass black holes which are not resolved in simulations with higher seed masses.  When considering the same mass scales, {\small{Astrid}} has fewer mergers, likely a result of the added infall time resulting from the dynamical friction model incorporated in {\small{Astrid}}.
    
    \item The merger rate for a given black hole is mass-dependent, with massive black holes undergoing mergers more frequently than low-mass black holes.  A typical black hole can be expected to undergo a merger every $\sim$1-10 Gyr.  
    
    \item Including low-mass black holes (i.e. lower mass seeding prescriptions) is crucial for modeling LISA detections.  High-mass seed models will only probe the low-frequency, high-strain regime within the LISA sensitivity band, while lower seed masses can extend across the full range of potential LISA detections.  
    
    \item In addition to decreasing the overall merger rate, including dynamical friction can be expected to preferentially affect low-frequency, low-strain GW signals, which are primarily generated by low-$M_2$ mergers.  This further emphasizes the importance of including accurate models for both seeding and infall dynamics, as both have the potential to not only affect the expected signal rate, but also the frequency-strain distribution of GW signals that LISA will be expected to detect.

    \item The SNR distribution follows a rough power law until the highest signals (SNR $> 10^2$), at which point there is a peak caused by $M_{\rm{BH}} \sim 10^6 M_\odot$ mergers.  This peak is poorly resolved in simulations with high seed masses, but is 0.5-1.5 dex above the {\small{Astrid}} seed masses, providing a well resolved sample.  Additionally, SNR is strongly correlated with the more massive black hole ($M_1$), and weakly correlated to the less massive black hole ($M_2$; though note that $M_1$ and $M_2$ are correlated to each other).  

    \item The LISA SNR has a weak dependence on redshift, at least in the early universe ($z > 2$).  There is evolution in the merging populations of black holes, with later times including higher-mass (and correspondingly smaller SNR) mergers.  However, low mass mergers have stronger signals, are more common, and have relatively weak dependence on redshift (at least for $z>3$, probed here).
    
\end{itemize}

In summary, we have shown that using a wide range of black hole seed masses extending down to $10^{4.5} \ h^{-1} \ M_\odot$ produces a high-mass population of black holes comparable to simulations which use higher seed masses, while simultaneously allowing us to investigate smaller black holes, mergers, and the associated gravitational waves they emit.  In particular, we are able to provide a detailed investigation into mergers spanning much of the LISA sensitivity range, provide estimates for the expected GW signal detections, and the correlation between these detections and the underlying merger masses and redshifts, in a simulation which directly models dynamical friction to produce more accurate merger behaviour.

\section*{Acknowledgements}
\small{Astrid} was run on the Frontera facility at the Texas Advanced Computing Center.  TDM acknowledges funding from the NSF AI Institute: Physics of the Future, NSF PHY-2020295, NASA ATP NNX17AK56G, and NASA ATP 80NSSC18K101.  TDM acknowledges additional support from NSF ACI-1614853, NSF AST-1616168, NASA ATP 19-ATP19-0084, and NASA ATP 80NSSC20K0519. SPB was supported by NASA ATP 80NSSC22K1897.

\section*{Data Availability}
The code used for the simulation is available at \url{https://github.com/MP-Gadget/MP-Gadget}.  Halo catalogs and BH data are available upon reasonable request to the authors.

\bibliographystyle{mn2e}       
\bibliography{astrobibl}       

\begin{appendix}
\section{Frequency Strain counts}

\begin{figure*}
    \centering
    \includegraphics[width=17cm]{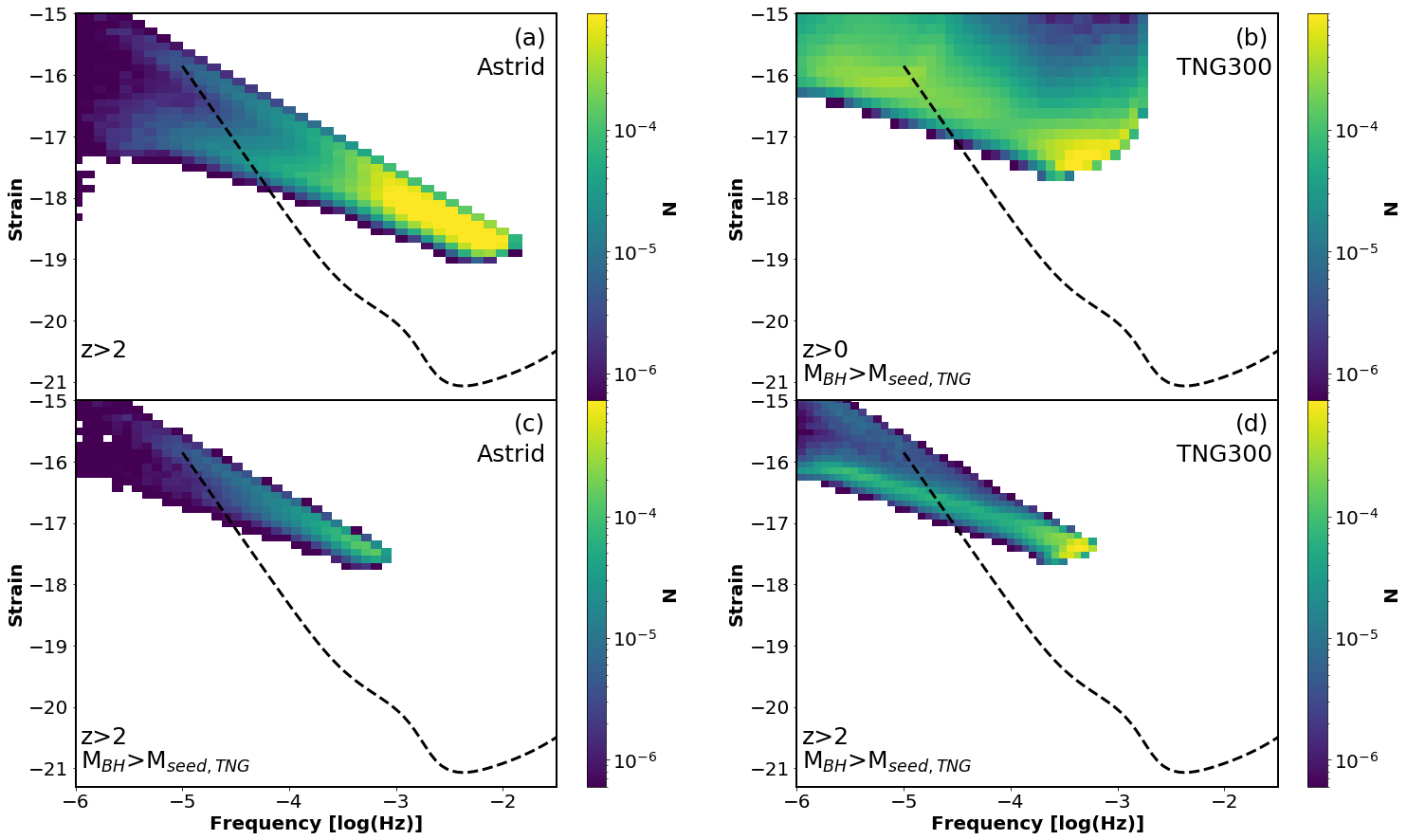}
    \caption{Same as Figure \ref{fig:freq_strain}, but colour coded by the number density of mergers, rather than the mean redshift.  We see that the merger frequency tends to peak toward higher-frequencies (corresponding to lower merger masses), and thus higher LISA SNR. This is strongest when considering lower mass mergers (e.g. when considering mergers down to the Astrid seed mass scale, in the top left panel), or when limited to high redshift (when black holes tend to be smaller).}
    \label{fig:freq_strain_colourcount}
\end{figure*}

Similar to Figure \ref{fig:freq_strain}, in Figure \ref{fig:freq_strain_colourcount} we show the frequency-strain distribution for both the Astrid and TNG300 simulations, but here color-coded by the number density of mergers.  Here we see that the peak in the frequency-strain distribution occurs at the highest frequencies, and lowest strain, corresponding to low-mass mergers taking  place at low redshift, consistent with the merger mass functions in Figure \ref{fig:massfunction_withmergers}.  Here we again see the importance of including lower-mass black hole seeds, as a higher seed mass will cut out the majority of mergers, which also correspond to the peak sensitivity range for LISA (compare Figure \ref{fig:freq_strain_colourcount} panel \textit{a} to panel \textit{c}).  By comparing Astrid to TNG, we also note that the higher seed mass results in a higher number of low-frequency GWs, especially at the lowest redshifts.

\end{appendix}

\end{document}